\documentclass[preprint,groupedaddress,amsmath,amssymb,prb]{revtex4}
\usepackage{longtable}
\usepackage{lscape}
\usepackage{fancyheadings}
\usepackage{amsmath}
\usepackage{amsfonts}
\usepackage{delarray}
\usepackage{graphicx}
\usepackage{dcolumn}
\usepackage{bm}

\newcommand {\apgt} {\ {\raise-.5ex\hbox{$\buildrel>\over\sim$}}\ }
\newcommand {\aplt} {\ {\raise-.5ex\hbox{$\buildrel<\over\sim$}}\ }

\DeclareMathSymbol{\blacklozenge}       {\mathord}{AMSa}{"07}

\begin{document}

\title{{First Principles Phase Diagram Calculations
for the Octahedral-Interstitial System {\bf $HfO_{X}$, 
$0 \leq X \leq 1/2$}} }

\author{Benjamin Paul Burton}
\email{benjamin.burton@nist.gov}
\affiliation{Materials Measyrement Laboratory,
Metallurgy Division, National Institute of Standards and Technology (NIST),
Gaithersburg, MD 20899, USA}
\altaffiliation{Phone: 301-975-6053, FAX: 301-975-5334.}

\author{Axel van de Walle}
\affiliation{Engineering and Applied Science Division,
California Institute of Technology,
1200 E. California Blvd.
MC 309-81
Pasadena, CA 91125;
avdw@alum.mit.edu}

\date{\today}

\begin{abstract}
First principles based phase diagram calculations were performed for the
hexagonal closest packed octahedral-interstitial solid solution 
system $\alpha HfO_{X}$~ 
($\alpha Hf[~~]_{1-X}O_{X}$; [~~]=Vacancy; $0 \leq X \leq 1/2$).
The cluster expansion method was used to do a ground state analysis,
and to calculate the phase diagram. The predicted diagram has four
ordered ground-states in the range $0 \leq X \leq 1/2$, but one of these,
at X=5/12, is predicted to disproportionate at T$\approx 220K$. 
At X$\approx 1/3$~ ($Hf_{3}O$) and X$\approx 1/2$~ ($Hf_{2}O$), 
order-parameter vs temperature plots evince a cascade of
ordered structures.


~~~\\
~~~\\
Submitted to {\bf Phys. Rev. B}

\end{abstract}

\maketitle


The group 4 hexagonal closest packed (hcp) suboxides
$MO_{X}$~ ($M~=~Ti,~Zr$~or $Hf$) all exhibit 
octahedral interstitial ordering of oxygen, $O$,
and vacancies, [~~], in solid solutions of the
form $\alpha M[~~]_{1-X}O_{X}$, $0 \leq X \leq 1/2$).
By far, the most studied of these systems is $ZrO_{X}$, 
because of issues related to the oxidation
of Zircalloy cladding on UO$_2$~ fuel rods in light-water reactors. 
\cite{Holmberg1961,Dubertret1966,Yamaguchi1968,Fehlmann1969,Yamaguchi1970,Hirabayashi1972,Hirabayashi1974,Hashimoto1974,Arai1976,Cronenberg1978,Hoffman1984a,Hoffman1984b,Hoffman1985,Sugizaki1985,Tsuji1995,Burton2011}
The hcp-based $HfO_{X}$~ system has attracted less
attention, \cite{Rudy1963,Hirabayashi1973,Okamoto1990,Shin2006} 
but Hafnium alloys are also potential
cladding materials; e.g. for long-lived nuclear waste transmutation 
applications in Boiling Water Reactors. \cite{Wallenius2008}
Also, investigating the chemical systematics of all
three group 4 suboxides enhances understanding of each
binary system.
In the $ZrO_{X}$ system, long-period superstructure (LPSS) phases 
were reported \cite{Fehlmann1969,Yamaguchi1970} in samples with
X$\approx$1/3, but not predicted in a recent
first principles phase diagram (FPPD) calculation \cite{Burton2011};
however, in the $HfO_{X}$ FPPD calculations described below, 
a cascade of related ordered structures is predicted at 
X$\approx$1/3 and X$\approx$1/2; it is not yet clear if this
cascade constitutes a Devil's Staircase. \cite{Bak1980,deFontaine1990}

\section{Methodology}

\subsection{Total Energy Calculations}

\begin{figure}[!htbp]
\begin{center}
\includegraphics[width=10.6cm,angle=0]{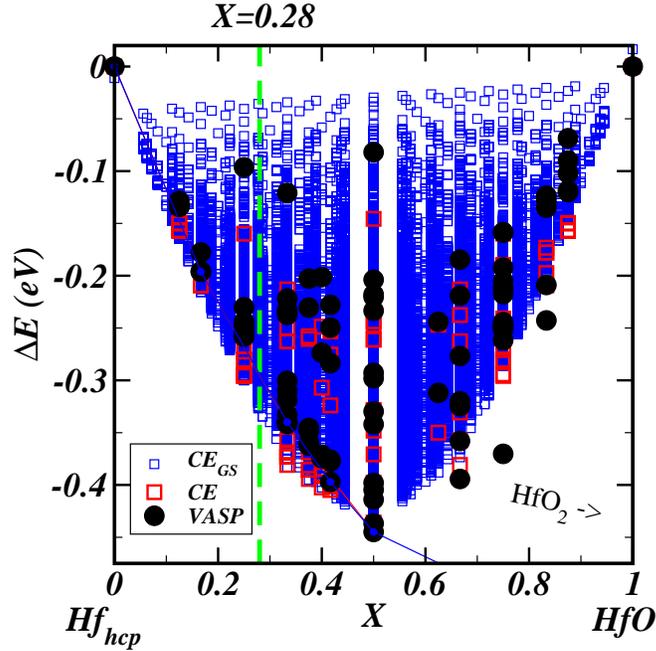}
\end{center}
\vspace{-0.5in}
\caption{Comparison of VASP (large solid circles) 
and CE (larger open squares, red online) 
formation energies,$\Delta E_{f}$, and a 
ground-state analysis on structures with 18 
or fewer octahedral-interstitial sites 
(smaller open squares, blue online). 
Extension of the convex hull towards the formation
energy of monoclinic Hafnia, $HfO_2$, 
indicates that the four ordered GS 
at X=1/6, 1/3, 5/12 and 1/2 are predicted
to be GS of the $Hf-O$~ binary. This is not in agreement
with experiment which suggests that the solubility
limit is approximately $HfO_{0.28}$~ (X$_{max} \approx 0.28$;
vertical dashed line, green online); 
i.e. in the solid solution $Hf_{1-Y}O_{Y}$, $Y \approx 0.22$.
} 
\label{fg:GS}
\end{figure}

Formation energies, $\Delta E_{f}$~ (Fig. \ref{fg:GS}) were calculated 
for fully relaxed hcp $\alpha Hf$, $HfO$ 
(hcp $\alpha Hf$ with all octahedral interstices occupied by $O$), and 96 
$\alpha Hf[~~]_{1-n}O_{n}$~ supercells of intermediate composition.
All calculations were performed with the density functional
theory (DFT) based Vienna $ab~initio$~
simulation program (VASP, version 4.4.5 \cite{Disclaimer,Kresse1993}) using 
projector-augmented plane-wave pseudopotentials, and the generalized gradient
approximation for exchange and correlation energies. 
Electronic degrees of freedom were optimized with a conjugate gradient
algorithm, and both cell constant and ionic positions were fully relaxed.

Total energy calculations were converged with respect
to k-point meshes by increasing the density of 
k-points for each structure until convergence is achieved.
A 500 eV energy cutoff was used, in the ``high precision" 
option which guarantees that {\it absolute}~ energies are 
converged to within a few meV/site 
(a few tenths of a kJ/site of exchangeable species; $O$, [~~]).  
Residual forces were typically 0.02 eV or less.

Calculated formation energies, $\Delta E_{f}$, 
relative to a mechanical
mixture of $\alpha Hf$ + $\alpha HfO$, for the 106 
$\alpha Hf[~~]_{1-n}O_{n}$~
supercells are plotted as solid circles in Fig. \ref{fg:GS}. 
Values of $\Delta E_{f}$~ are, 

\begin{equation}
\Delta E_{f} = (E_{Str} - E_{\alpha Hf} - E_{\alpha HfO} )/(2)
\end{equation}

\noindent
where: $E_{Str}$~ is the total energy of the 
$\alpha Hf[~~]_{1-n}O_{n}$~ supercell; 
$E_{\alpha Hf}$ is the energy/atom of $\alpha Hf$;
$E_{\alpha HfO}$ is the energy/atom of $\alpha HfO$.

\subsection{The Cluster Expansion Hamiltonian}

The cluster expansion, CE \cite{Sanchez1984}, is a compact representation 
of the configurational total energy. In the 
$\alpha Hf[~~]_{1-X}O_{X}$ system, 
the solid solution configuration is described by pseudospin occupation
variables $\sigma _{i}$, which take values $\sigma_i=-1$~ when site-$i$ is 
occupied by [~~] and $\sigma_i=+1$~ when site-$i$ is occupied by $O$.

The CE parameterizes the configurational energy, per exchangeable cation, as
a polynomial in pseudospin occupation variables:

\begin{eqnarray}
E(\mathbf{\sigma })=\sum_{\ell }m_{\ell }J_{\ell }\left\langle
\prod_{i\in \ell ^{\prime }}\sigma _{i}\right\rangle
\end{eqnarray}

\noindent
Cluster $\ell$~ is defined as a set of lattice sites. The sum is taken over all
clusters $\ell$~ that are not symmetrically equivalent in the
high-T structure space group, and  the average is taken 
over all clusters $\ell ^{\prime }$ that are
symmetrically equivalent to $\ell$.
Coefficients $J_{\ell}$~ are called effective cluster interactions, ECI, and
the \emph{multiplicity} of a cluster, $m_{\ell}$, is the
number of symmetrically equivalent clusters, divided by the number of cation sites.
The ECI are obtained by fitting a set of VASP FP calculated 
structure energies, $\{ E_{Str} \}$.
The resulting CE can be improved as necessary by increasing the
number of clusters $\ell$~ and/or the number of $E_{Str}$~ used in the fit.

\begin{figure}
\begin{center}
\vspace{-0.5in}
\includegraphics[width=8.0cm,angle=0]{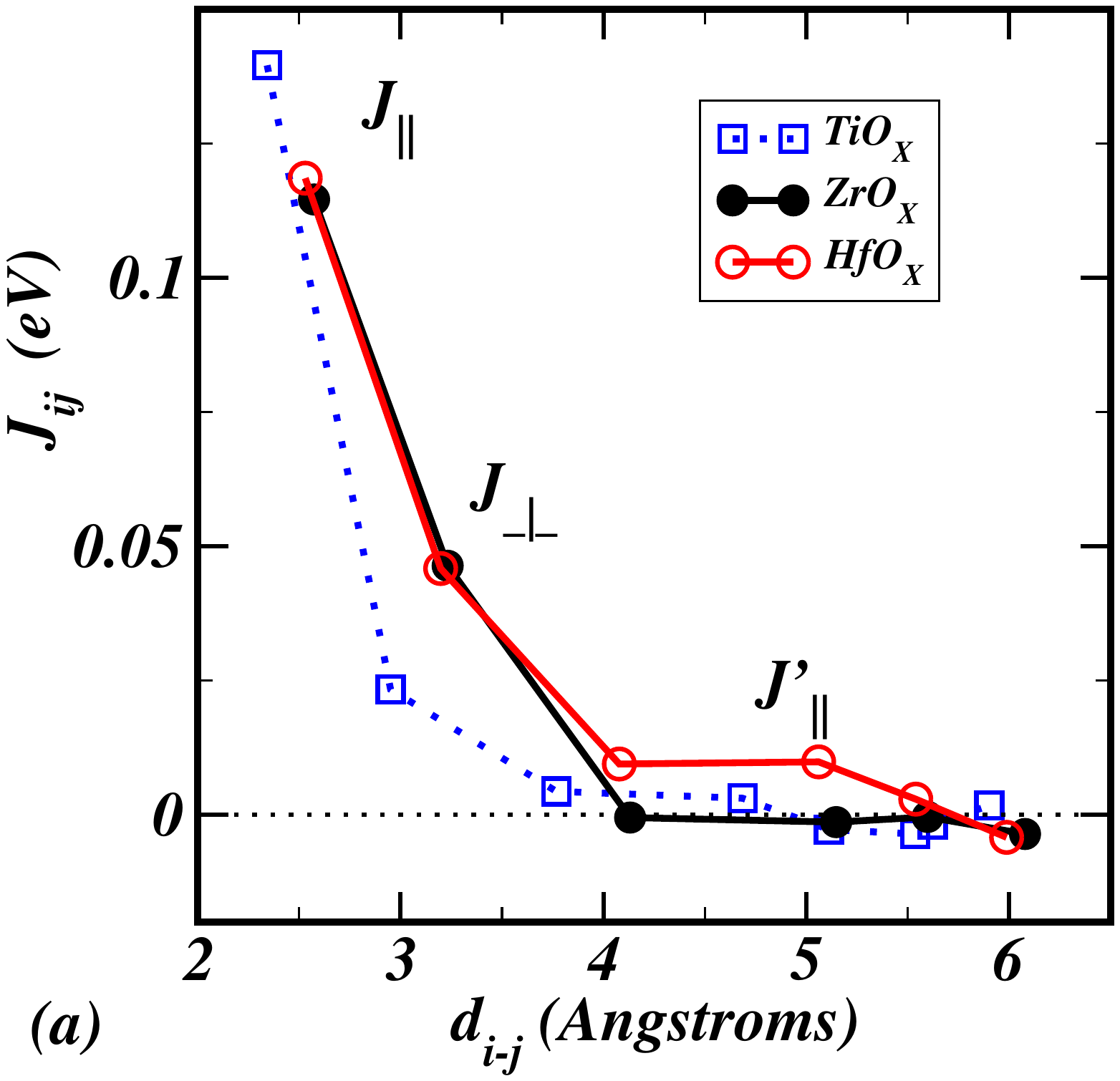}
\hspace{-0.3in}
\includegraphics[width=8.0cm,angle=0]{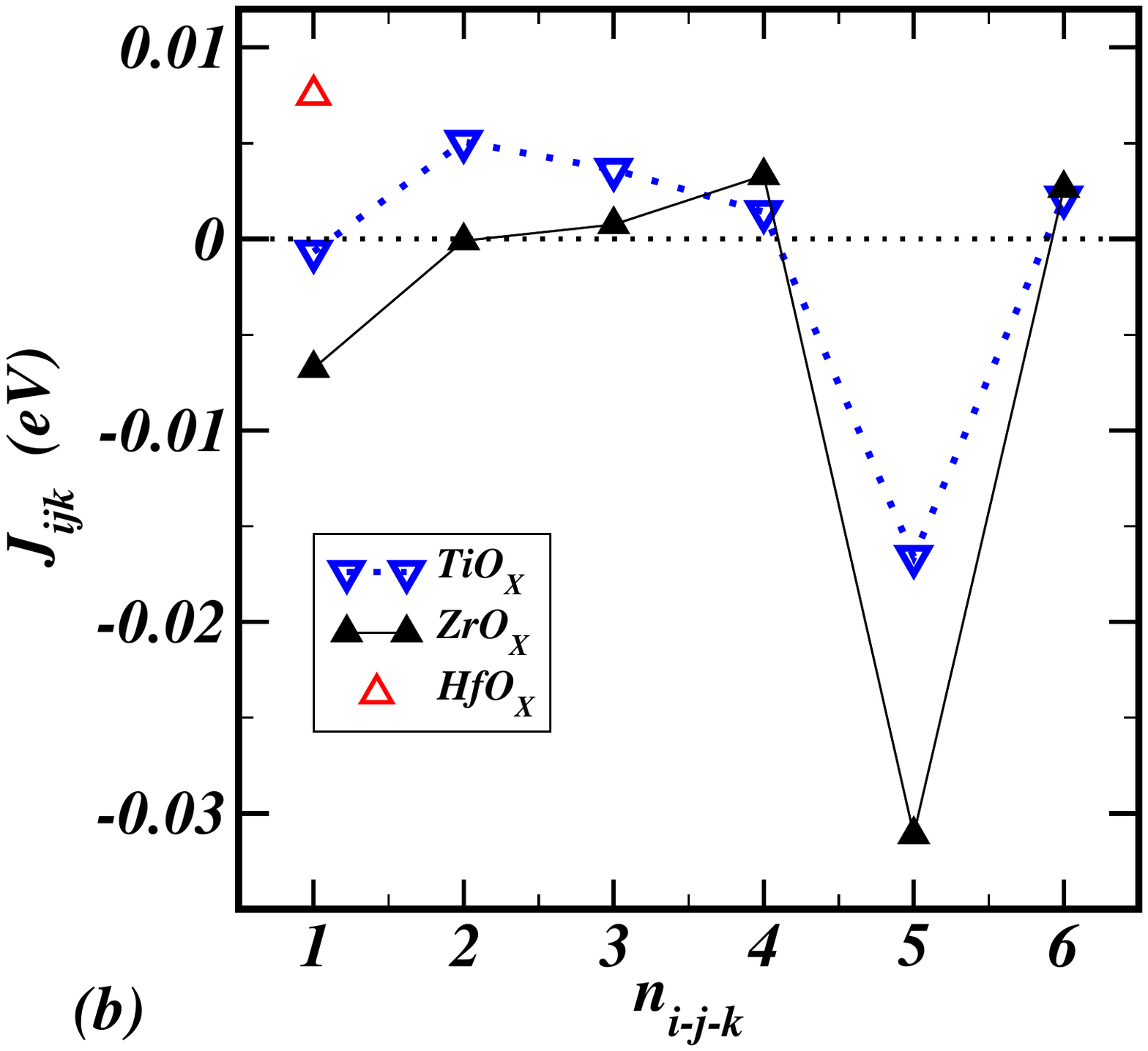}
\end{center}
\caption{Effective Cluster Interactions (ECI) for pair and
3-body interactions.  
Solid and dotted lines are to guide the eye. Results
for the systems $TiO_{X}$ (open blue symbols online), 
$ZrO_{X}$ (solid black symbols) and 
$HfO_{X}$ (open red symbols online).
(a) The first two pair-ECI are for
nearest-neighbor (nn) $O$-[~~] pairs that are 
parallel- ($J_{\parallel}$) and perpendicular ($J_{\perp}$), 
respectively, to the hexagonal c-axis; the second pair parallel
to c$_{Hex}$ is $J^{\prime}_{\parallel}$~ (pairwise-ECI are 
plotted as functions of inter-site separation), note that the
results for $HfO_{X}$ are very similar to those for
$ZrO_{X}$~ except that the 3'rd and 4'th ($J^{\prime}_{\parallel}$)
nn-pairs are significantly larger in $HfO_{X}$;
(b) 3-body interactions are plotted as functions of the
index n$_{i-j-k}$ ~ which increases, nonlinearly,
as the area of triangle i-j-k increases. Large positive pairwise
ECI imply strong pairwise $O$-[~~] nn-attractions,
i.e. strong pairwise $O-O$~ nn-repulsions.}
\label{fg:eci}
\end{figure}

Fitting was performed with the Alloy Theoretic Automated Toolkit (ATAT)
\cite{Axel2002a,Axel2002b,Axel2002c,Disclaimer} 
which automates most of the tasks associated with the 
construction of a CE Hamiltonian.  A complete description
of the algorithms underlying the code can be found in
\cite{Axel2002b}. The zero- and point-cluster values were -0.571537 eV 
and 0.013973 eV, respectively.  The six pair and one 3-body ECI are 
plotted in Figs. \ref{fg:eci}a and \ref{fg:eci}b (open symbols, red online),
as are ECI for $ZrO_{X}$ (solid black symbols) and $TiO_{X}$ 
(open symbols, blue online).  As in $ZrO_{X}$ and $TiO_{X}$, 
nearest neighbor (nn) $O-O$ pairs are highly energetic, 
and therefore strongly avoided; hence nn-pair 
ECI are strongly $attractive$~ (ECI $\verb+>+0$, for $O$-[~~] nn pairs); but 
beyond nn-pairs, the pairwise ECI are smaller; however the 3'rd and 
4'th nn pair-ECI in $HfO_{X}$~ are significantly larger than 
corresponding terms for $ZrO_{X}$~ and $TiO_{X}$. As in $ZrO_{X}$,
the ratio of ECI parallel ($J_{\parallel}$) and perpendicular
($J_{\perp}$) to c$_{Hex}$, respectively, is 
$J_{\parallel} / J_{\perp} \approx 2.5$; for $TiO_{X}$, 
$J_{\parallel} / J_{\perp} \approx 5$.  
These results are similar to those presented by Ruban et al. \cite{Ruban2010},
although the ECI presented here are not identically comparable owing to
different treatments of relaxation energies.

\section{Results and Discussion}

\subsection{Ground-States}

The CE was used for a ground-state (GS) analysis that
included all configurations of [~~] and $O$~ in systems of
18 or fewer $Hf$-atoms (octahedral interstitial sites); 
a total of $2^{18} = 262,144$~ structures (reduced by symmetry).
Five GS were identified in the range, $0 \leq X \leq 1/2$, 
i.e. at X = 0, 1/6, 1/3, 5/12 and 1/2;
solid circles (blue online) on the convex hull (solid line) 
in Fig. \ref{fg:GS}.
The extension of the convex hull towards monoclinic
hafnia ($HfO_2$) is also plotted in Fig. \ref{fg:GS}. The
CE-results suggest that all four VASP-GS in the 
$\alpha Hf[~~]_{1-X}O_{X}$ subsystem are also GS of the $Hf-O$~ binary.
The VASP-predicted maximum solubility of $O$~ in $Hf$~ is 
X$_{max} \approx 0.5$, significantly greater than the
experimental value of X$_{max} \approx 0.28$. 
 
Larger open squares (red online) in Figure \ref{fg:GS} are
CE-calculated values for the $\Delta E_{f}$~ that correspond 
to the VASP calculations, and the smaller open squares (blue online)
are $\Delta E_{f}$~ for the remaining 262,144-106=262,038 structures
in the GS analysis.  All space group determinations
were performed with the FINDSYM program. \cite{Disclaimer,FINDSYM}

\pagebreak

\begin{longtable}[!t]{|c|c|c|c|l|}
\caption[]{Crystal structure parameters for 
predicted ground-state phases in the $\alpha Hf[~~]_{1-X}O_{X}$~ 
system.  Cell constants are given in $\AA$.} \\ \hline
System       &     X      & Space Group       &Calculated cell            &Idealized              \\
             &  atomic    & IT number         & constants                 &Atomic                 \\
             & fraction $O$~ & Pearson Symbol    & ($\AA$)                   & Coordinates           \\ \hline \hline
$Hf_6O$      &   1/6      & R$\overline{3}$   &$a \approx \surd \overline{3} a_0$& $O$:  0,  0,  0   \\ 
             &            & 148               &$ = 5.5391$                & $Hf$: 1/3,  0, 5/12   \\
             &   1/7      & hP7               &$c \approx 3c_0 = 15.183$  & $Hf$:  0   , 1/3 , 5/12 \\ 
             &            &                   &                           & $Hf$: 2/3, 2/3, 5/12    \\
             &            &                   &                           & $Hf$: 2/3,  0, 7/12  \\
             &            &                   &                           & $Hf$:  0,  2/3, 7/12   \\
             &            &                   &                           & $Hf$: 1/3, 1/3, 7/12    \\ \hline \hline
$Hf_3O$      &   1/3      & P$\overline{3}$1c &$a \approx \surd \overline{3} a_0$& O: 1/3, 2/3, 1/4  \\
             &            & 163               &$ = 5.5391 $               &  $O$:   2/3, 1/3, 3/4  \\
             &   1/4      & hP16              &$c \approx 2c_0 = 10.122$  &  $O$:    0,   0,   0   \\ 
             &            &                   &                           &  $O$:    0,   0,  1/2  \\ 
             &            &                   &                           & $Hf$:   2/3, 2/3, 7/8  \\ 
             &            &                   &                           & $Hf$:   1/3,  0,  7/8  \\ 
             &            &                   &                           & $Hf$:    0,  1/3, 7/8  \\ 
             &            &                   &                           & $Hf$:    0,  2/3, 5/8  \\ 
             &            &                   &                           & $Hf$:   1/3, 1/3, 5/8  \\ 
             &            &                   &                           & $Hf$:   2/3,  0,  5/8  \\ 
             &            &                   &                           & $Hf$:    0,  1/3, 3/8  \\ 
             &            &                   &                           & $Hf$:   2/3  2/3, 3/8  \\ 
             &            &                   &                           & $Hf$:   1/3   0,  3/8  \\ 
             &            &                   &                           & $Hf$:   1/3  1/3, 1/8  \\ 
             &            &                   &                           & $Hf$:   2/3   0,  1/8  \\ 
             &            &                   &                           & $Hf$:    0,  2/3, 1/8  \\ \hline \hline
$Hf_{12}O_5$ &  5/12    & R$\overline{3}$   &$a \approx \surd \overline{3} a_0$& $O$:  0,    0,   1/12  \\
             &            & 148               & $ = 10.615 $              & $O$: 0,   0,  11/12     \\
             &  5/17      & hP17              & $c \approx 6c_0 = 30.366$ & $O$: 0,   0,  1/3        \\ 
             &            &                   &                           & $O$: 0,   0,  2/3         \\ 
             &            &                   &                           & $O$: 0,   0,  1/2         \\ 
             &            &                   &                           & $Hf$: 2/3, 2/3,  13/24    \\
             &            &                   &                           & $Hf$: 1/3,  0,   13/24    \\
             &            &                   &                           & $Hf$:  0,  1/3,  13/24    \\
             &            &                   &                           & $Hf$: 1/3, 1/3,  11/24    \\
             &            &                   &                           & $Hf$: 2/3,  0,   11/24    \\
             &            &                   &                           & $Hf$:  0,  2/3,  11/24    \\
             &            &                   &                           & $Hf$: 2/3, 2/3,  3/8    \\
             &            &                   &                           & $Hf$: 1/3,  0,   3/8    \\
             &            &                   &                           & $Hf$:  0,  1/3,  3/8    \\
             &            &                   &                           & $Hf$: 1/3, 1/3,  5/8    \\
             &            &                   &                           & $Hf$: 2/3,  0,   5/8    \\
             &            &                   &                           & $Hf$:  0,  2/3,  5/8    \\ \hline \hline
$Hf_2O$      &   1/2      & P$\overline{3}$1m &$a \approx \surd \overline{3} a_0$& $O$: 0, 0, 0     \\
             &            & 162               & $ = 5.5391 $              & $O$:  1/3, 2/3, 1/2     \\
             &   1/3      & hP9               &$c \approx c_0 = 5.0610$   & $O$:  2/3, 1/3, 1/2     \\
             &            &                   &                           & $Hf$: 1/3,  0,  1/4     \\
             &            &                   &                           & $Hf$:  0,  1/3, 1/4     \\ 
             &            &                   &                           & $Hf$: 2/3, 2/3, 1/4     \\
             &            &                   &                           & $Hf$: 2/3,  0,  3/4     \\
             &            &                   &                           & $Hf$:  0,  2/3, 3/4     \\
             &            &                   &                           & $Hf$: 1/3, 1/3, 3/4     \\ \hline \hline
\end{longtable}

\pagebreak

\begin{figure}[!htbp]
\vspace{-1.0in}
\hspace{-1.0in}
\begin{center}
\includegraphics[width=20.cm,angle=0]{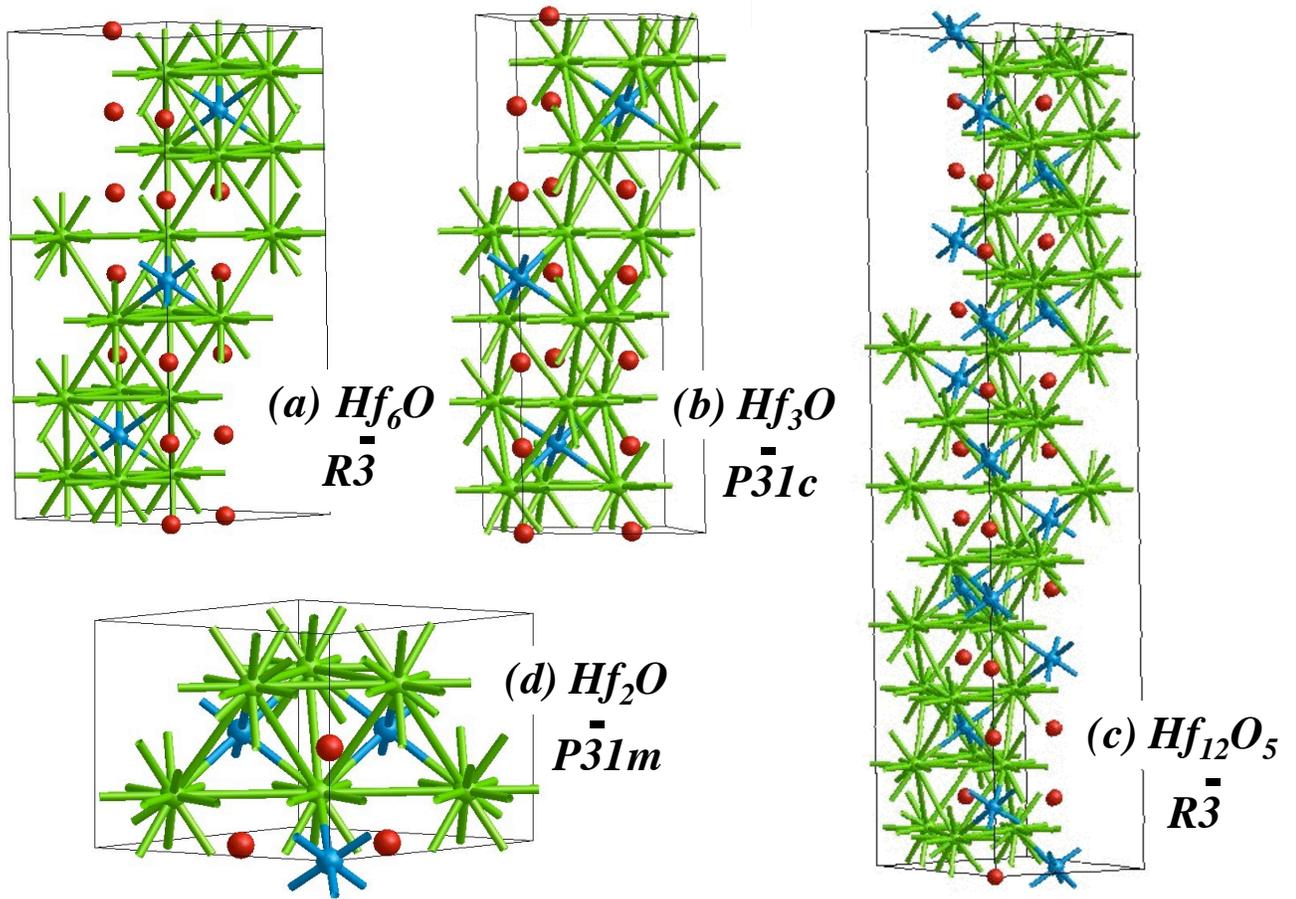}
\end{center}
\vspace{-0.5in}
\caption{Idealized crystal structures of the four
cluster-expansion-predicted suboxide ground-states: 
(a) $Hf_6O$; (b) $Hf_3O$; (c) $Hf_{12}O_5$; (d) $Hf_2O$. 
Spheres connected by bond-sticks (yellowish-green
online) represent $Hf$. Isolated spheres with bond-sticks
(blue online) represent oxygen. Isolated spheres
(red online) represent vacant octahedral sites.
} 
\label{fg:XalGS}
\end{figure}
\pagebreak

Ground State crystal structures of the VASP- and CE-GS in $Hf-HfO$~ are 
described in Table~I and their idealized structures 
are drawn in Figures \ref{fg:XalGS} a-d:  
where $Hf$~ is represented by spheres connected with bond-sticks 
(yellowish-green online); $O$~ is represented by isolated 
spheres with bond-sticks (blue online); and [~~] are 
represented by isolated spheres (red online).  As in the
$ZrO_{X}$ system, all GS structures are characterized by
$O-O$~ nn-avoidance both parallel- and perpendicular to c$_{Hex}$.  

The VASP-CE-predicted R$\overline{3}$~ $Hf_{6}O$~ GS is
the same as the experimental low-T structure reported by Hirabayashi et al.
\cite{Hirabayashi1973}. Space group relations, 
require a first-order P6$_{3}$mmc $\leftrightharpoons$ R$\overline{3}$~ 
disorder $\leftrightharpoons$ order transition between the P6$_{3}$mmc disordered
phase and the R$\overline{3}$~ $Hf_{6}O$~ ordered phase. The $Hf_{6}O$~ GS is the 
only GS within the experimental solubility range $0 \leq X \aplt 0.28$; 
all the other computationally predicted GS-phases are presumably 
metastable.

\subsection{Finite Temperature Calculations}

\subsubsection{The Phase Diagram}

\begin{figure}[!htbp]
\begin{center}
\includegraphics[width=12.0cm,angle=0]{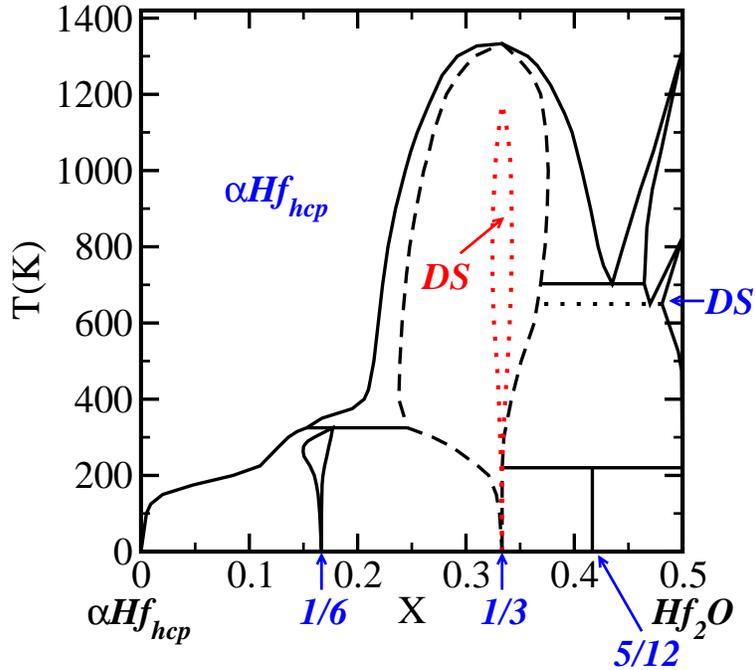}
\end{center}
\caption{Calculated
phase diagram for the system $\alpha Hf[~~]_{1-X}O_{X}$:
Approximate regions in which the calculations
predict a cascade of ordered phases is labeled DS? to
indicate possible Devil's Staircases of closely related 
ordered phases.
}
\label{fg:XT}
\end{figure}

A first principles phase diagram (FPPD) calculation was performed
with conoical- and grand canonical Monte Carlo (MC) simulations using the 
emc2 code which is part of the ATAT
package \cite{Axel2002a,Axel2002b,Axel2002c}. Input parameters for
emc2 were: a simulation box with at least 4,050 octahedral sites; 
2000 Monte Carlo passes. 
The predicted phase diagram is shown in Figure \ref{fg:XT}.
Most phase boundaries were determined by following order-parameters
($\eta$) of the various ordered phases as functions of X and T.
Dotted boundaries are used to acknowledge uncertainties
in phase boundary determinations.  In particular, boundaries
of the possible Devil's Staircase (DS? in Fig. \ref{fg:XT}) regions are 
are labeled DS?) are poorly defined, and the interior structures 
of these regions are undetermined.

\begin{figure}[!htbp]
\begin{center}
\vspace{-0.5in}
\includegraphics[width=8.cm,angle=0]{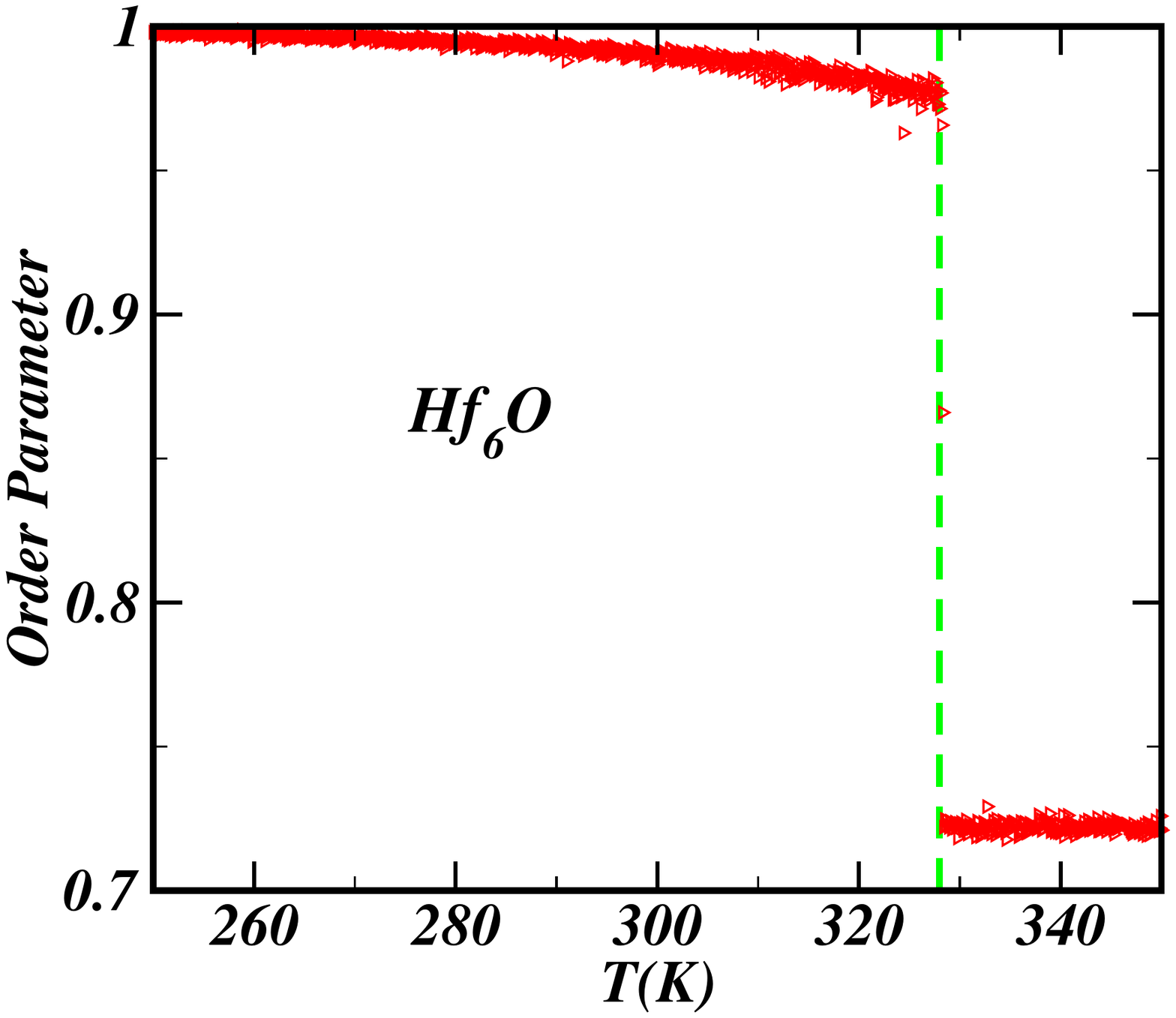}
\end{center}
\caption{Calculated order-parameter vs temperature (T) curve
for the calculated 1'st order transition in
$Hf_{6}O$.
}
\label{fg:Hf6O}
\end{figure}

\subsubsection{$Hf_6O$}

Interstitial ordering of $O$~ and [~~] in hcp $HfO_{X}$~ was
studied by Hirabayashi et al. \cite{Hirabayashi1973}
who used electron- and neutron diffraction to analyse single 
crystals with bulk compositions in the range of 
$0 ~ \leq at \verb+%+ ~ O \aplt 20$~ ($0 \leq X \aplt 0.25$);
described as $HfO_{1/6-}$ and $HfO_{1/6+}$~ for samples with less than
or more than one $O$-atom per six $Hf$-atoms.
The structure that  Hirabayashi et. al. \cite{Hirabayashi1973}
report for $HfO_{1/6-}$~ has R$\overline{3}$~ space group symmetry and 
is identical to the VASP-GS at $Hf_{6}O$~ (Fig. \ref{fg:Hf6O} and 
Table~I).

The FPPD-predicted order-disorder transition in $Hf_{6}O$~ ($\approx 325K$) 
is first-order (Fig. \cite{fg:Hf6O}), but significantly $lower$~ than than 
the experimental value ($\approx 700K$ \cite{Hirabayashi1973}) or the 
calculated value from Ruban et al. (600K; \cite{Ruban2010} predicted 
transition-order not reported).  Typically, FPPD calculations $overestimate$~ ]
order-disorder transition temperatures so this result is surprising. 

\begin{figure}[!htbp]
\begin{center}
\vspace{-0.7in}
\includegraphics[width=8.cm,angle=0]{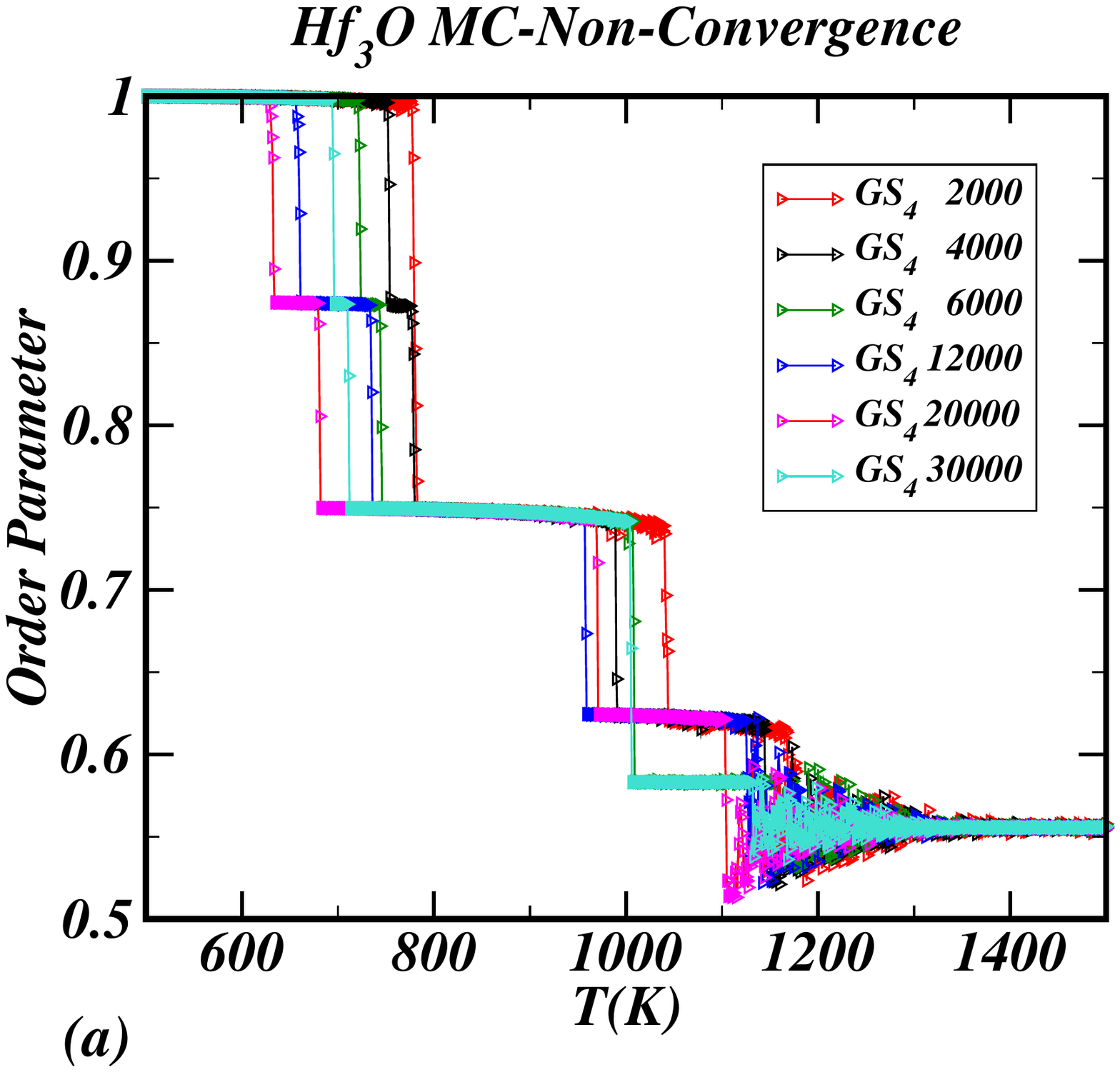}
\hspace{-0.3in}
\includegraphics[width=8.cm,angle=0]{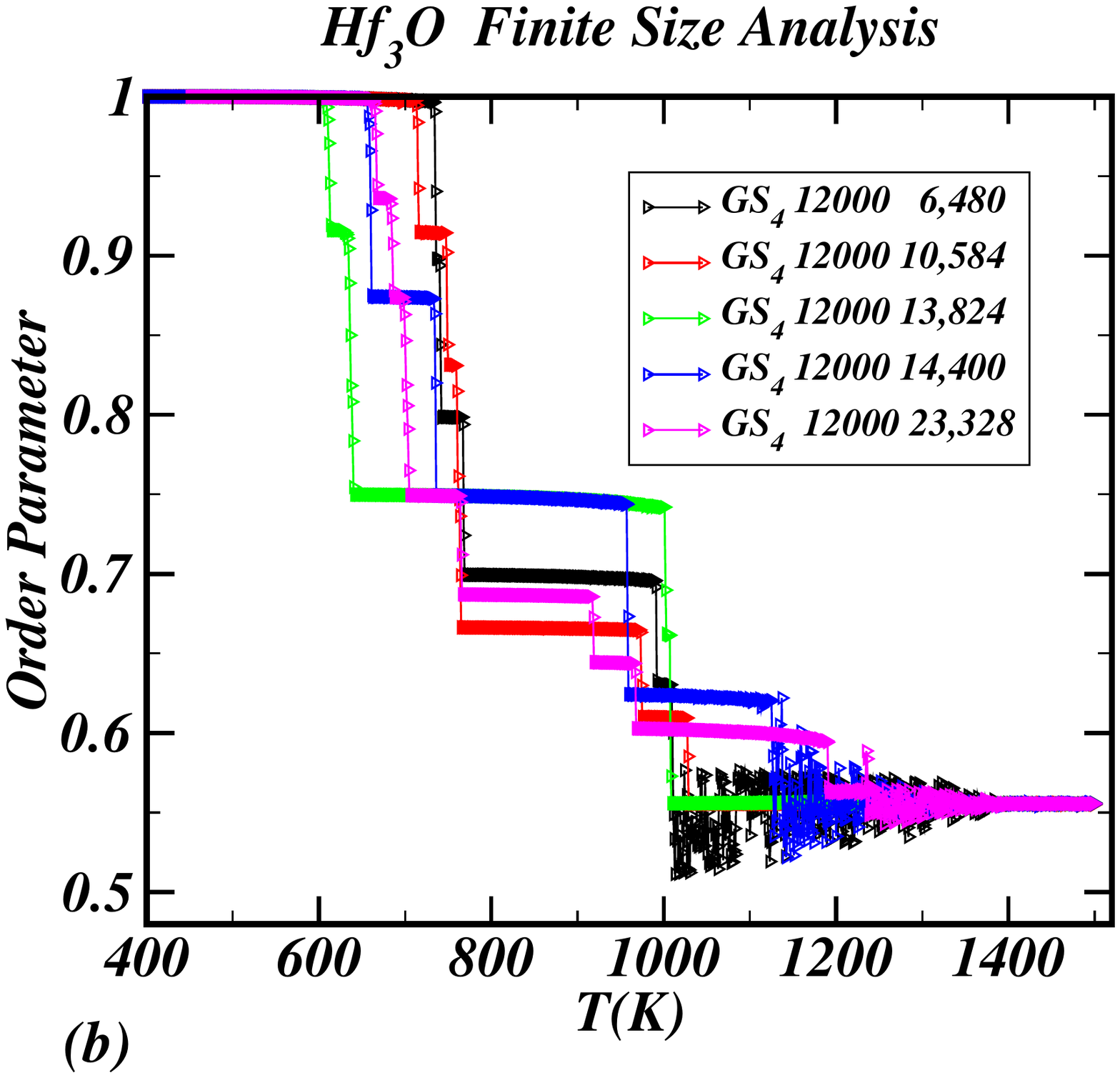} \\
\vspace{-1.3in}
\includegraphics[width=8.cm,angle=0]{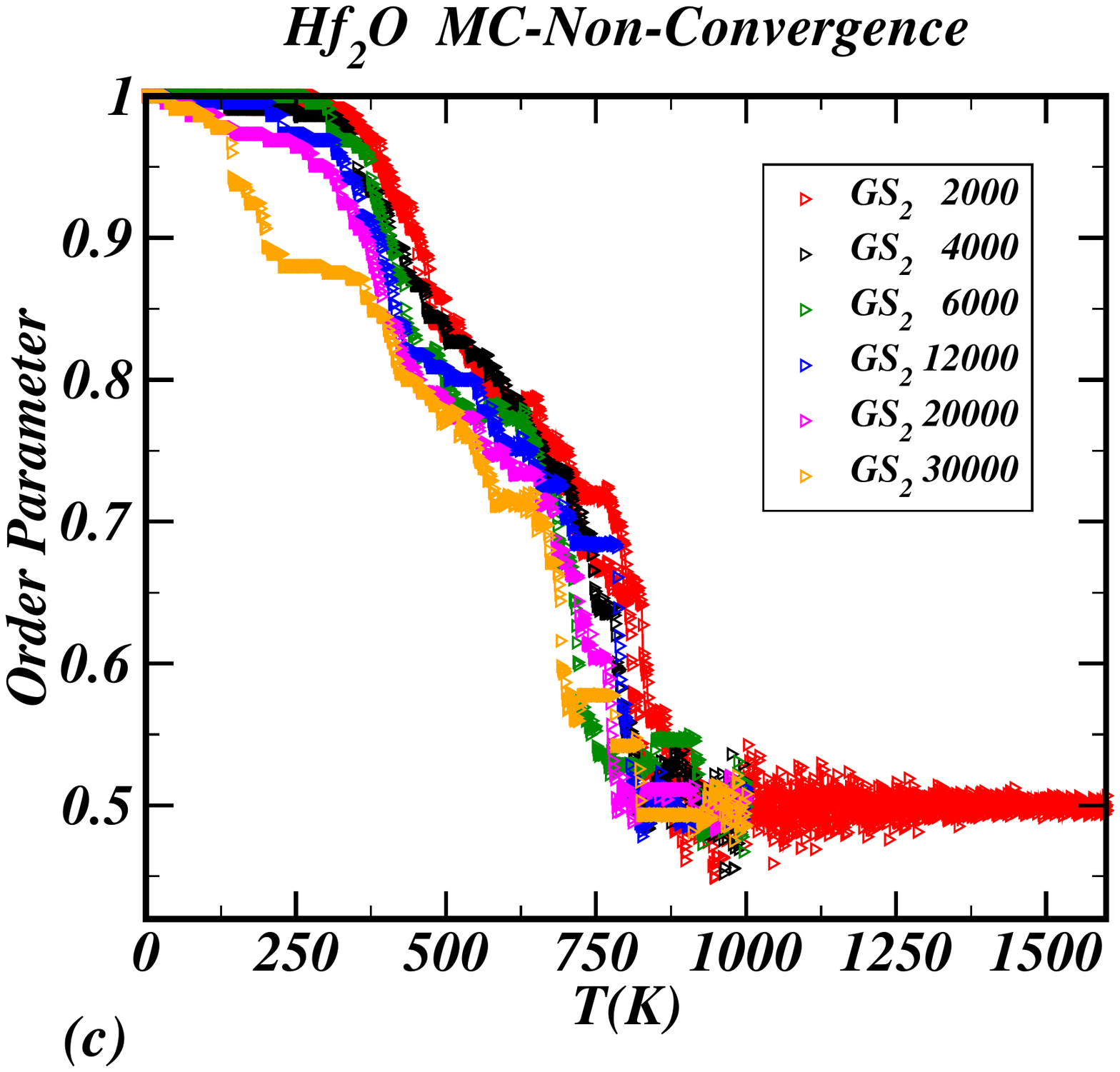}
\hspace{-0.3in}
\includegraphics[width=8.cm,angle=0]{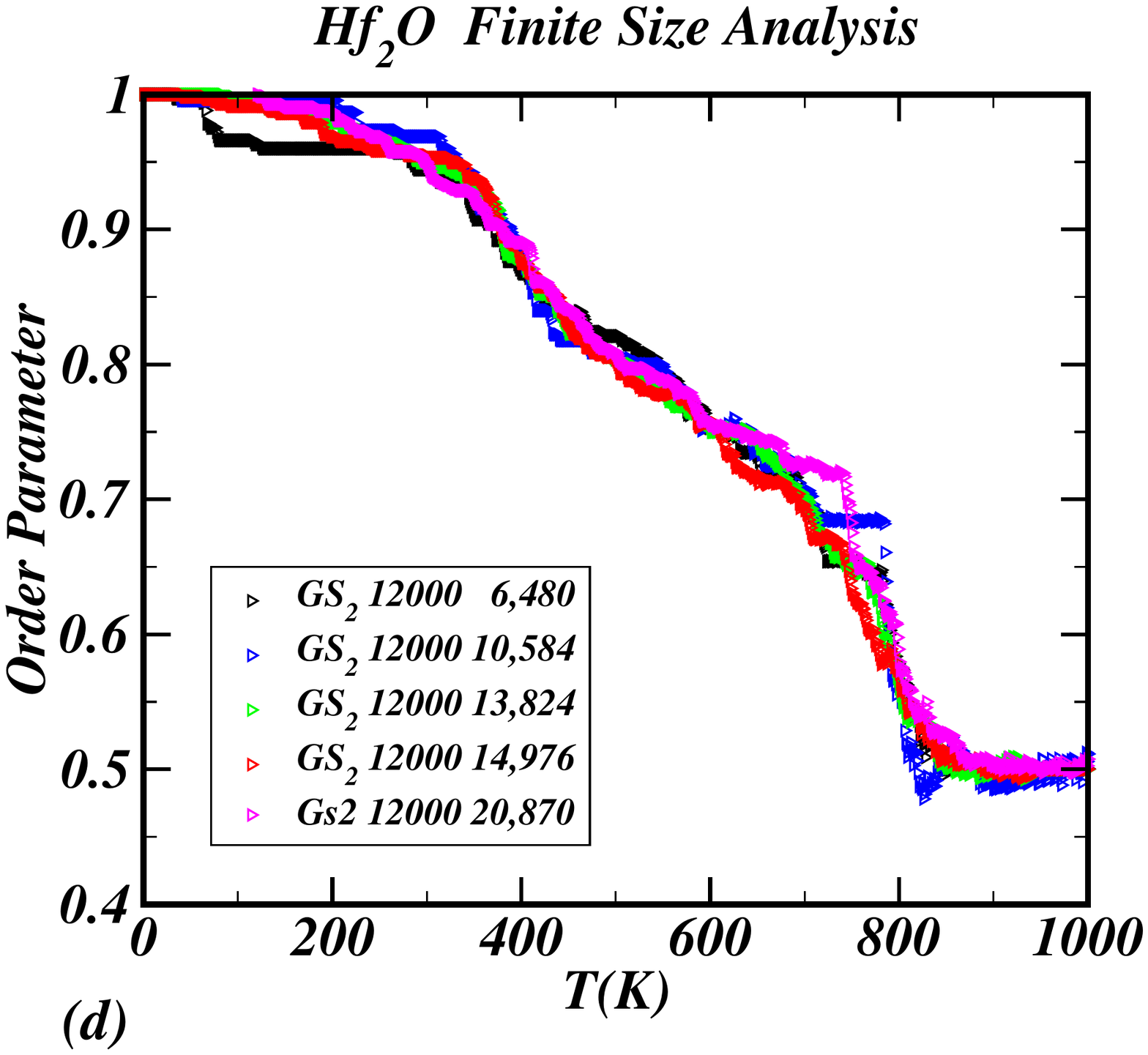}
\end{center}
\vspace{-0.5in}
\caption{Calculated order-parameter vs temperature (T) curves
that evince possible Devil's Staircases (DS?) of closely
related ordered phases at X=1/3 (a,b) and X=1/2 (c,d):
(a,c) Monte-Carlo (MC) simulations at constant MC-box-size 
and different numbers of MC-passes 
  at each T, 
almost always find the same set of ordered-phase plateaus
(the notation GS$_{4}$ 2000 means the simulation was 
started in the $Hf_{3}O$-GS and run for 2000 MC-passes per T);
(b,d) Calculations at a constant number of MC-passes, with
various MC-box-sizes yield different plateau sequences because 
box-size determines allowed periodicities for ordered phases 
(the notation GS$_{4}$ 12,000 6,480 means the simulation was
started in the $Hf_{3}O$-GS and run for 12,000 MC-passes per T,
on an MC-box that contains 6,480 sites for $O$:[~~] mixing).
}
\label{fg:DS}
\end{figure}

\begin{figure}[!htbp]
\begin{center}
\vspace{-0.8in}
\includegraphics[width=16.cm,angle=0]{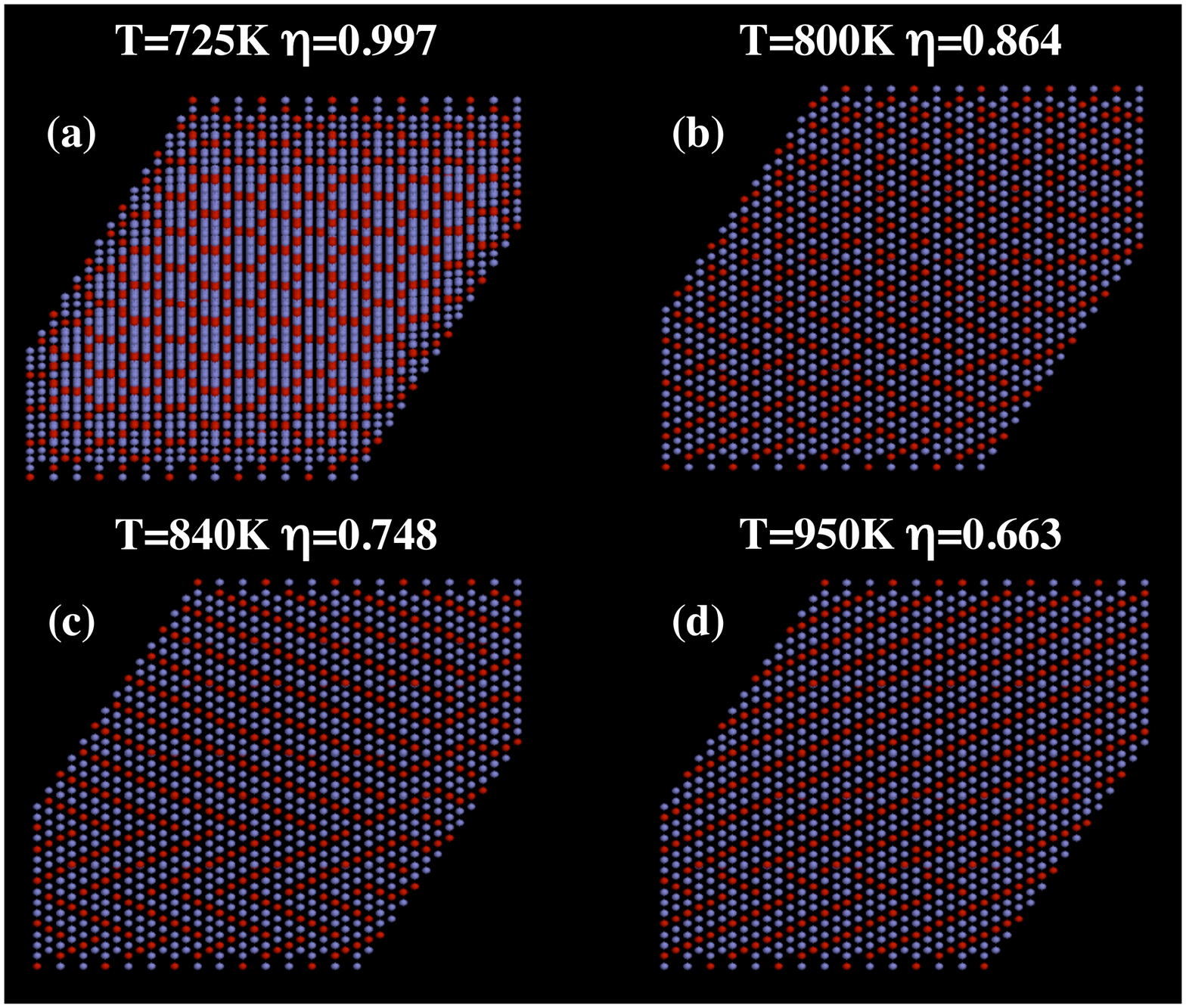} \\
\vspace{-1.in}
\hspace{0.048in}
\includegraphics[width=16.cm,angle=0]{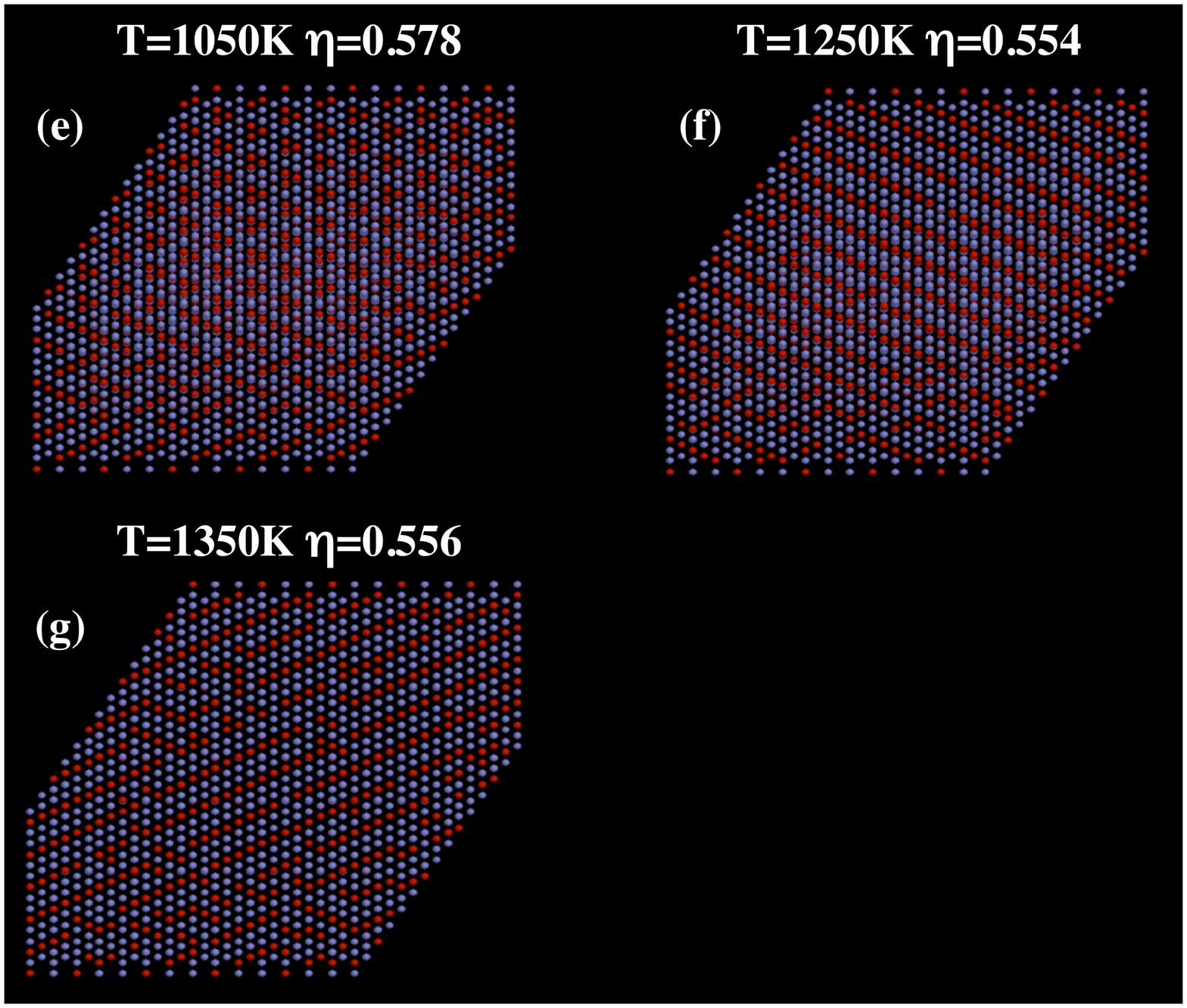} 
\end{center}
\vspace{-0.51in}
\caption{MC snapshots of ordered phases in the possible 
Devil's Staircase in $Hf_3O$. For clarity, only 
O- (red online) and vacant-sites ([~~]-sites, blue online)
are shown: (a) The GS-structure with minor disorder.
In (b)-(g) there appears to be a competition
between two ordering modes: 1) a striped mode (c,d,f,g) 
in which layers ${\perp}$~ to c$_{Hex}$~ have single O-rows 
that alternate with i double-[~~]-rows; 
2) a triangular mode (b,e) in which layers ${\perp}$~ to c$_{Hex}$~ 
exhibit an ordered array of $O_3$-nn-equalateral triangles and
[~~]$_6$-nn-equalateral triangles.
}
\label{fg:Snap}
\end{figure}

\subsubsection{Possible Devil's Staircase in $Hf_3O$}

The most $O$-rich structure determination in 
Hirabayashi et al.  \cite{Hirabayashi1973} was for 
a sample with X=0.203.  
The reported structure has 
P$\overline{3}$1c symmetry, and is equivalent to the
predicted $Hf_{3}O$~ structure 
[Fig. \ref{fg:XalGS}(b), Table I] 
except that in the experimental sample, maximal 
$O$-site occupancy would be $ \aplt ([~~]_{0.16},O_{0.84})$.
 
Figures \ref{fg:DS}a and b are plots of order-parameter vs. T for 
for $Hf_3O$. The results plotted in Fig. \ref{fg:DS}a, were
calculated with the MC-box-size held constant at 4,050 
$O$:[~~]-sites while the number of MC-passes is varried.  
Almost all the order-parameter plateaus
are the same for different numbers of MC-passes, which reflects
the influence of MC-box size on the ordered-phase periodicities 
that are allowed. Note that the transition temperatures from one 
plateau to another are clearly not converged.
The results plotted in Fig. \ref{fg:DS}b were calculated with
a constant numbers of MC-passes (12,000) and various MC-box-sizes; 
i.e. $O$:[~~]-sites.  Varying MC-box-size allows different ordered-phase
periodicities; i.e. allows access to more stairs in the cascade
of ordered phases.

In Fig. \ref{fg:Snap}a-g Monte Carlo snapshots are shown at seven different 
temperatures. For clarity, only O- (red online) and [~~]-sites (blue online)
are shown.  These MC-snapshots appear to indicate a competition between
two ordering modes: 1) a striped mode (c,d,f,g)
in which layers ${\perp}$~ to c$_{Hex}$~ exhibit single O-rows 
that alternate with double-[~~]-rows;
2) a triangular mode (b,e) in which layers ${\perp}$~ to c$_{Hex}$~ exhibit
ordered arrays of $O_3$-nn-equalateral triangles and
[~~]$_6$-nn-equalateral triangles. It is not clear if this
cascade of ordered structures constitutes a Devil's Staircase,
but the results presented in Fig. \ref{fg:DS}a and b suggest that
it does.

\subsubsection{Possible Devil's Staircase in $Hf_2O$}

In Figures \ref{fg:DS}c and d, one sees similar order-parameter 
vs. T systematics for $Hf_2O$~ as one finds in 
Figures \ref{fg:DS}a and b 
for $Hf_3O$; except that the density of plateaus is much greater 
in $Hf_2O$, because a 1:1 $O$:[~~]-ratio allows a greater number
different periodic ordered structures than a 2:1 $O$:[~~]-ratio. 

\section{Conclusions}

Ground-State ordered phases are predicted at 
X=0, 1/6, 1/3, 5/12 and 1/2, but only those
at X=0 or X=1/6 are likely to be physically
realized because the experimental value for
the maximum solubility of $O$~ in hcp $HfO_X$~
is $X_{max} \approx 0.28$.  

Observed ordered phases at X=1/6 and X=0.203 \cite{Hirabayashi1973}
agree with predicted GS at X=1/6 and X=1/3 (but with
diluted $O$-site occupancies).

In the metastable portion of the $HfO_X$~ phase diagram,
(0.28 ~ \apgt ~X) cascades of ordered phases, possible Devil's Staircases, 
are predicted for bulk compositions near $Hf_3O$~ and $Hf_2O$.



\begin{thebibliography}{999}
\bibitem{Disclaimer}
Disclaimer: the use of a specific software package should
not be misinterpreted as implying a NIST endorsement of 
that package.
\end{thebibliography}
\end{document}